\documentclass[twocolumn, aps, prl,amsmath,amssymb,citeautoscript,longbibliography]{revtex4-2}

\pdfoutput=1
\usepackage{natbib}
\usepackage[english]{babel}
\usepackage{subfigure}

\usepackage{letltxmacro}
\usepackage{latexsym}
\LetLtxMacro{\ORIGselectlanguage}{\selectlanguage}
\makeatletter
\DeclareRobustCommand{\selectlanguage}[1]{%
  \@ifundefined{alias@\string#1}
    {\ORIGselectlanguage{#1}}
    {\begingroup\edef\x{\endgroup
       \noexpand\ORIGselectlanguage{\@nameuse{alias@#1}}}\x}%
}
\newcommand{\definelanguagealias}[2]{%
  \@namedef{alias@#1}{#2}%
}
\makeatother
\definelanguagealias{en}{english}
\definelanguagealias{English}{english}
\usepackage{graphicx}
\usepackage{amsmath}
\usepackage{amsfonts}
\usepackage{amssymb,bbding}
\usepackage{bm}
\usepackage[title]{appendix}
\usepackage{color}
\usepackage[percent]{overpic}
\usepackage{soul} 
\usepackage{amssymb}
\usepackage{wasysym}
\usepackage{dsfont}
\usepackage{float}
\usepackage{braket}
\usepackage{cancel}
\usepackage{comment}
\usepackage[utf8]{inputenc}

\usepackage{enumitem}
\usepackage[dvipsnames]{xcolor}
\usepackage{graphicx}

\usepackage{mathtools}
\usepackage{comment}
\usepackage{relsize}
\usepackage{dsfont} 
\usepackage{commath}
\usepackage{todonotes}
\usepackage{titlesec}
\usepackage{hyperref}
\hypersetup{
	colorlinks=true,
	linkcolor=blue,
	filecolor=magenta,      
	urlcolor=cyan,
	pdftitle={Emergence of Gaussianity in the thermodynamic limit of interacting fermions}
}

\usepackage{bm}
\usepackage{cleveref}

\setlist[itemize]{leftmargin=*}
\setlist[enumerate]{leftmargin=*}

\crefformat{equation}{Eq.~(#2#1#3)}
\Crefformat{equation}{Eq.~(#2#1#3)}
\crefrangeformat{equation}{Eqs.~(#3#1#4--#5#2#6)}
\crefname{figure}{Fig.}{Fig.}

\DeclareRobustCommand*{\ora}{\overrightarrow}

\begin{document}

\title{Emergence of Gaussianity in the thermodynamic limit of interacting fermions}

\author{Gabriel Matos}
\author{Andrew Hallam}
\author{Aydin Deger}
\author{Zlatko Papi\'c}
\author{Jiannis K. Pachos}
\affiliation{School of Physics and Astronomy, University of Leeds, Leeds LS2 9JT, United Kingdom}

\date{\today}

\begin{abstract}

Systems of interacting fermions can give rise to ground states whose correlations become effectively free-fermion-like in the thermodynamic limit, as shown by Baxter for a class of integrable models that include the one-dimensional  XYZ spin-$\frac{1}{2}$ chain. Here, we quantitatively analyse this behaviour by establishing the relation between system size and correlation length required for the fermionic Gaussianity to emerge. Importantly, we demonstrate that this behaviour can be observed through the applicability of Wick's theorem and thus it is experimentally accessible. 
To establish the relevance of our results to possible experimental realisations of XYZ-related models, we demonstrate that the emergent Gaussianity is insensitive to weak variations in the range of interactions, coupling inhomogeneities and local random potentials.

\end{abstract}

\maketitle{}
	
{\bf \em Introduction:--} Free-particle systems enjoy a privileged place in physics: all of their correlations can be broken down into products of two-point functions, illustrating the computational power of Wick's theorem~\cite{Peschel2009} and greatly aiding their theoretical understanding. This ``Gaussian" description can be radically altered in real systems where  interactions are invariably present, leading to exotic interaction-driven phenomena such as fractionalised excitations and topological order~\cite{AKLT,Laughlin1983}. At the same time, there are many known examples,  e.g., Luttinger liquids~\cite{Giamarchi2004}, where interactions give rise to new collective degrees of freedom, however the latter can still be described as nearly free. It is thus important to have a more systematic understanding of the criteria when interactions can engender nontrivial physical behaviour. 

In recent years significant attention has been focused on many-body systems that are expected to be strongly interacting yet behave in an approximately  Gaussian manner. Recent experiments~\cite{schweigler2021decay} have shown that Gausssian behaviour can emerge \emph{dynamically} as the system is taken out of its equilibrium state. On the other hand, Gaussianity can also emerge in \emph{equilibrium}, as the \emph{size} of the system grows infinite. The latter occurs in a one-dimensional (1D) spin-1/2 XYZ model, which hosts a variety of paradigmatic models, such as the Heisenberg model, the XY-model and the Ising model,  as special cases. In 1970, Sutherland observed that the transfer matrix of the eight-vertex model has the same eigenvectors as the XYZ model~\cite{Sutherland1970}. With the help of this mapping Baxter famously solved the 2D classical XYZ model exactly. In particular, he demonstrated that in the thermodynamic limit its partition function can be described by non-interacting fermions throughout its entire phase diagram~\cite{Baxter1971b,PhysRevLett.26.832,Baxter1985, baxter1973eight}. With the mapping between 2D classical systems and quantum chains~\cite{Nishino1995, Nishino1997,takhtadzhyan1979quantum} it was possible to determine the entanglement spectra of the 1D spin-$1/2$ XYZ model directly from Baxter's results~\cite{Peschel1999, Peschel2009}. Subsequently, the Gaussian structure of the entanglement spectra of the XYZ chain has been verified numerically~\cite{Roy2021}. 

In this paper, we propose to use the violation of Wick's theorem for measuring the fermionic Gaussianity emerging in the XYZ model and its generalisations to long-range spin-spin interactions. When applied to a free fermion system, Wick's theorem decomposes high-point correlators in terms of two-point correlators \cite{Peskin:1995ev}. When the system is interacting, this decomposition is not possible, leaving a difference, ${\cal W}$, that can be used to quantify the effect of interactions. This approach makes it possible to demonstrate the emergent freedom of a many-body quantum system using simple, physical observables. However, ${\cal W}$  is dependent on the choice of the operator used for Wick's decomposition and therefore needs an upper bound to be physically useful.  We demonstrate the efficacy of Wick's theorem by bounding it with the more general diagnostic of Gaussianity -- the so-called \emph{interaction distance}, $D_{\cal F}$~\cite{turner_optimal_2017,pachos_quantifying_2018,patrick_interaction_2019}. 
Comparing the behaviour of both $D_{\cal F}$ and ${\cal W}$, we demonstrate that quantum correlations of the XYZ model exponentially approach those of a free-fermion model as a function of system size, provided the size of the system is larger than the correlation length. For smaller system sizes, the XYZ model appears to be strongly interacting both in terms of Wick's theorem and $D_\mathcal{F}$ near its critical regions, in stark contrast to its thermodynamic limit behaviour.  To demonstrate the experimental relevance of our results, we analyse the applicability of Wick's in the presence of realistic conditions such as variations in the range of interactions, coupling inhomogeneities and local random potentials.

{\bf \em The XYZ model and its emergent freedom.--} The 1D spin-$1/2$ XYZ model on an open chain with $L$ sites is given by 
\begin{equation} \label{eq:xyz_model}
  H = \sum_{i} J_x X_i X_{i+1} + J_y Y_i Y_{i+1} + J_z Z_i Z_{i+1},
\end{equation}
where $X_{i}$, $Y_{i}$, $Z_{i}$ are the usual Pauli matrices on site $i$. By employing the Jordan-Wigner transformation, the spin model maps to interacting spinless fermions
\begin{equation} 
\label{eq:fermion_model}
  H = \sum_{i} J_+ c_i c^\dagger_{i+1} + J_{-} c_i c_{i+1} + \text{h.c.} + \frac{J_z}{4} n_i n_{i+1}  -\frac{J_z}{2}  n_i,
\end{equation}
where $J_{\pm} =(J_x\pm J_y)$ and $n_i=c^\dagger_i c_i$. In the fermionic representation $J_z$ becomes the interaction coupling between fermion populations at neighbouring sites~\cite{LiebSchultzMattis}. Without loss of generality we take $J_x{=}1$ and due to the symmetries $(J_y, J_z) \leftrightarrow (-J_z, -J_y)$, $(J_y, J_z) \leftrightarrow (J_z, J_y)$ of the Hamiltonian, we restrict ourselves to $J_y{\geq} 0$. 

\begin{figure}[tb]
	\centering
	\includegraphics[width=0.9\linewidth]{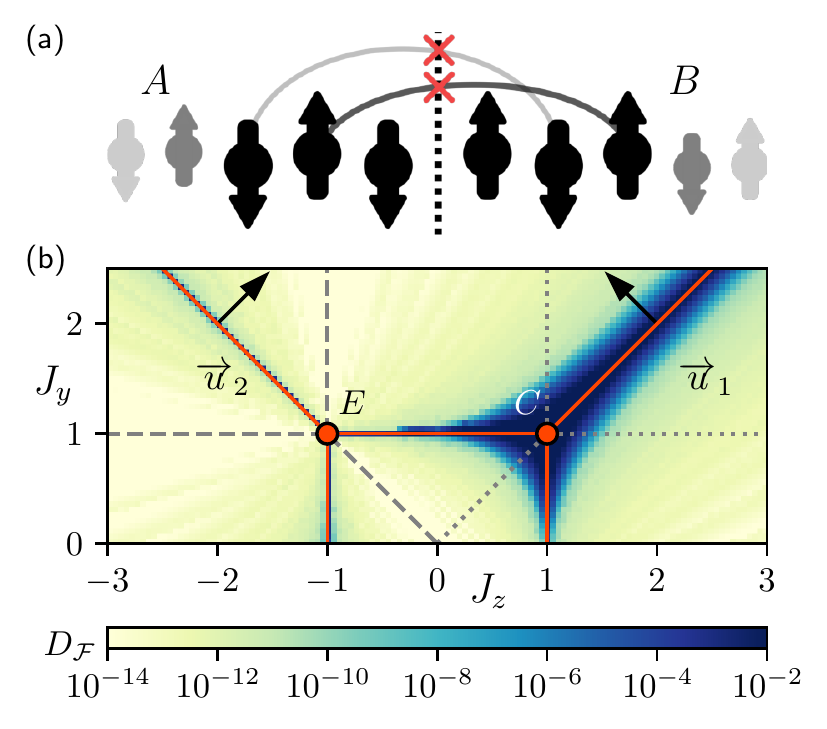}
	\caption{(a) Quantum correlations across the bipartition (dashed line) of a spin chain. (b) Interaction distance, $D_\mathcal{F}$, in Eq.~(\ref{eq:df}), obtained using DMRG across the phase diagram of the XYZ model for $L{=}200$ spins. Red lines denote critical lines, and the conformal ($C$) and non-conformal ($E$) tricritical points are indicated~\cite{ercolessi2011essential}. Vectors $\ora{u}_1, \ora{u}_2$  are orthogonal to the critical lines and are used in Fig.~\ref{fig:xyz_scalings}. Interaction distance is strongly suppressed  in gapped phases of the XYZ model, signalling the emergence of Gaussianity.}
	\label{fig:xyz_df_vne_pd}
\end{figure}

\begin{figure}[t]
\centering
\includegraphics[width=0.8\linewidth]{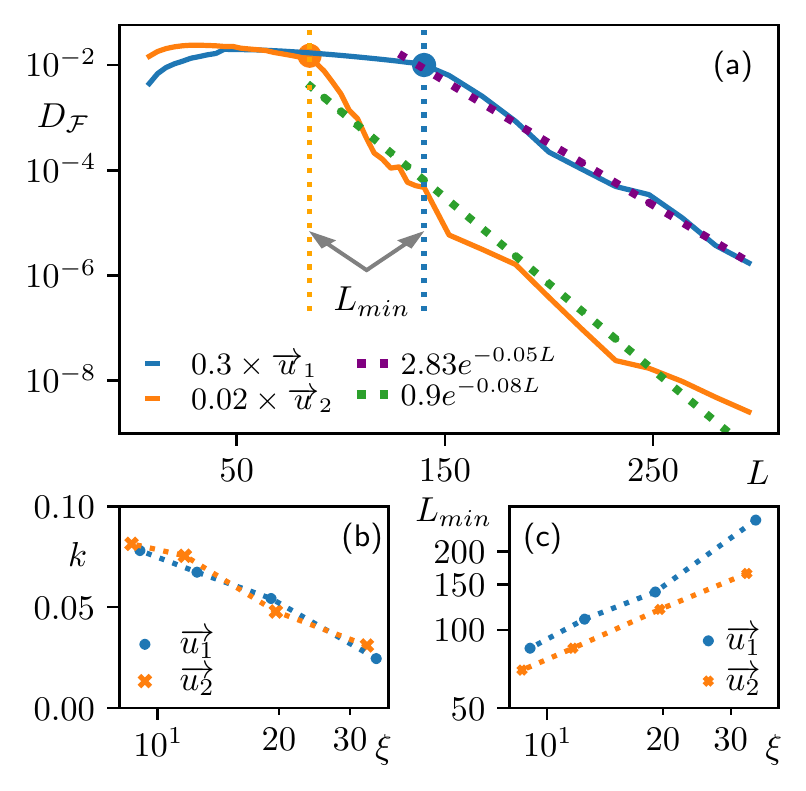}
\caption{(a) Exponential decay of $D_\mathcal{F}$ with system size at different points along $\ora{u_1}$ and $\ora{u_2}$ cuts through the phase diagram in Fig.~\ref{fig:xyz_df_vne_pd}. We observe a short initial increase, folllowed by a plateau and the final decrease beyond some crossover lengthscale, $L_\mathrm{min}$, indicated by the dotted lines. 
	Dashed lines are fits to the asymptotic exponential decay, $D_{\cal{F}} \propto \exp(-k L)$, for data points $L{>}L_\mathrm{min}$.
	(b) Slope $k$ of the exponential decay, extracted at various points along $\ora{u_1}$ and $\ora{u_2}$,  exhibits a power-law dependence on correlation length $\xi$. The latter is computed using the analytic formulas applicable in the thermodynamic limit~\cite{SOM}.  (c) Correlation length $\xi$ displays power-law dependence on $L_\mathrm{min}$.
}
\label{fig:xyz_scalings}
\end{figure}

To analyse quantum correlations in the ground-state of the model $\ket{\psi_\text{GS}}$, we take the reduced-density matrix of the half-chain, $\rho_A {=} \text{tr}_B \ket{\psi_\text{GS}}\!\bra{\psi_\text{GS}}$, illustrated in Fig.~\ref{fig:xyz_df_vne_pd}(a). The eigenvalues $\rho_i$ of $\rho_A$, and the corresponding entanglement energies, $E^\text{ent}_i {=} -\ln \rho_i$, contain all information about quantum correlations between the two halves of the chain~\cite{li_entanglement_2008}. The total amount of correlations can be quantified by the von Neumann entropy $S(\rho_A) {=} {-}\sum_i \rho_i\ln \rho_i$. On the other hand, the interaction distance, $D_{\cal F}(\rho_A)$, diagnoses how close the quantum correlations between the two halves are to those of a Gaussian fermionic state~\cite{turner_optimal_2017}. The interaction distance is defined as
\begin{equation}
\label{eq:df}
D_{\cal F} (\rho_A) = \min_{\{ \bm{ \epsilon} \}}{1 \over 2} \sum_i | \rho_i - \sigma_i(\bm{\epsilon})|,
\end{equation}
where $\sigma_i(\bm{\epsilon})$ are the eigenvalues of the density matrix $\sigma(\bm{\epsilon})$ of a free model given by $\sigma_i(\bm{\epsilon}) {=} e^{-E_i^\text{free}(\bm{\epsilon})}$ with entanglement spectrum $E_i^\text{free}(\bm{\epsilon}) = E_0 +\sum_j \epsilon_j^{} n_j^{(i)}$, where $E_0$ ensures the normalisation condition $\text{tr} (\sigma){=} 1$, $\epsilon_j$ are the single particle energies and $n_j^{(i)}$ is the occupancy number on the $j$th site of the $i$th element of the Fock basis, labelled by the index $i$. The minimisation over the single particle energies $\bm{\epsilon}$ guaranties that $\sigma$ is the free density matrix which is closest to the interacting $\rho$~\cite{markham_quantum_2008}. When $D_{\cal F}{\to} 0$ then the ground state of the model exhibits Gaussian correlations across the bipartition and can be faithfully described by a free-fermion density matrix  $\sigma$.

The interaction distance for the XYZ model in Eq.~(\ref{eq:xyz_model}), shown in Fig.~\ref{fig:xyz_df_vne_pd}(b), is computed across the phase diagram using density matrix renormalisation group (DMRG)~\cite{WhiteDMRG}, implemented in iTensor~\cite{itensor}.  Along the line $J_y {=} J_z$, which is equivalent to XXZ model with antiferromagnetic couplings studied in Ref.~\onlinecite{patrick_interaction_2019}, $D_\mathcal{F}$ is high around the gapless critical phase $\abs{J_y} {>} 1$. On the line $J_y {=}{-} J_z$, $D_\mathcal{F}$ is high across a much narrower region around its $\abs{J_y} {>} 1$ gapless phase. Away from the critical regions, $D_{\cal F}$ tends to zero showing that the system exhibits Gaussian correlations. 
Hence, we will focus our investigation around these gapless regions where $D_\mathcal{F}$ exhibits non-trivial behaviour. 

Baxter proved that the XYZ model becomes free when $L{\to}\infty$,  provided the correlation length is finite. However, physically, we expect the model becomes free as soon as $L$ exceeds the correlation length, $\xi$. 
When applied to the XYZ model, the interaction distance can diagnose the emergence of freedom and thus quantify Baxter's result for various system sizes $L$ compared to  $\xi$. Without loss of generality, we consider the behaviour of $D_{\cal F}$ along $\ora{u_1}$ cut across the $J_y{=}J_z$ critical region and $\ora{u_2}$ that crosses the $J_y{=}{-}J_z$ critical region, as shown in Fig.~\ref{fig:xyz_scalings}(b).  We find that, for values of the couplings away from the critical lines, $D_{\cal F}$ tends to zero exponentially fast as system size increases, as shown in Fig.\ref{fig:xyz_scalings}(a), signalling that the emerging freedom can be observed efficiently. We emphasise that this happens even for large values of the coupling $J_z$ that correspond to strong density-density interactions between fermions. 

To analyse the conditions under which the system becomes free, we determine the system size $L_\mathrm{min}$ beyond which the interaction distance starts decreasing as well as the rate $k$ at which $D_{\cal F}$ exponentially approaches zero, $D_\mathcal{F}{\propto}\exp(-k L)$. Fig.~\ref{fig:xyz_scalings}(b) shows that the rate $k$ decreases as correlation length $\xi$ increases, i.e. the rate of exponential decay of $D_{\cal F}$ decreases the closer we are to the critical regions.  Hence, $D_{\cal F}$ quantifies Baxter's assumption, showing the quantum correlations of the XYZ model become free-fermion-like by having $D_{\cal F}{\to} 0$ exponentially fast with $L$, provided that the size is larger than a minimum value $L_\mathrm{min}$. The latter is a polynomial function of the correlation length $\xi$~\cite{SOM}, as can be seen in Fig.~\ref{fig:xyz_scalings}(c). We observe that the larger the correlation length, i.e., the closer to criticality, the larger the system needs to be in order for the interaction distance to exhibit the exponential decay. The observed polynomial relation between $L_\text{min}$ and $\xi$ quantifies Baxter's assumption for identifying the freedom of the XYZ model~\cite{Baxter1985}.  We emphasize that this strong dependence of $D_{\cal F}$ on $L$ allows to efficiently identify the emergent Gaussianity in a quantum simulation of the XYZ model with an exponential accuracy just with a linear cost in the size of the simulated system.

{\bf \em Violation of Wick's theorem and experimental implications.--} Investigating the behaviour of the XYZ model in terms of the interaction distance reveals its formal emergence of Gaussianity in a quantitative way. Ideally, we would like to have an experimentally accessible quantity that would allow us to measure the emergent freedom in the laboratory. In general, the full entanglement spectrum of the system can be difficult to extract in an experimental context \cite{pichler2016measurement, dalmonte2018quantum, Kokail2021}. We therefore turn to the violation of Wick's theorem due to interactions. 

\begin{figure}[t]
  \centering
  \includegraphics[width=0.9\linewidth]{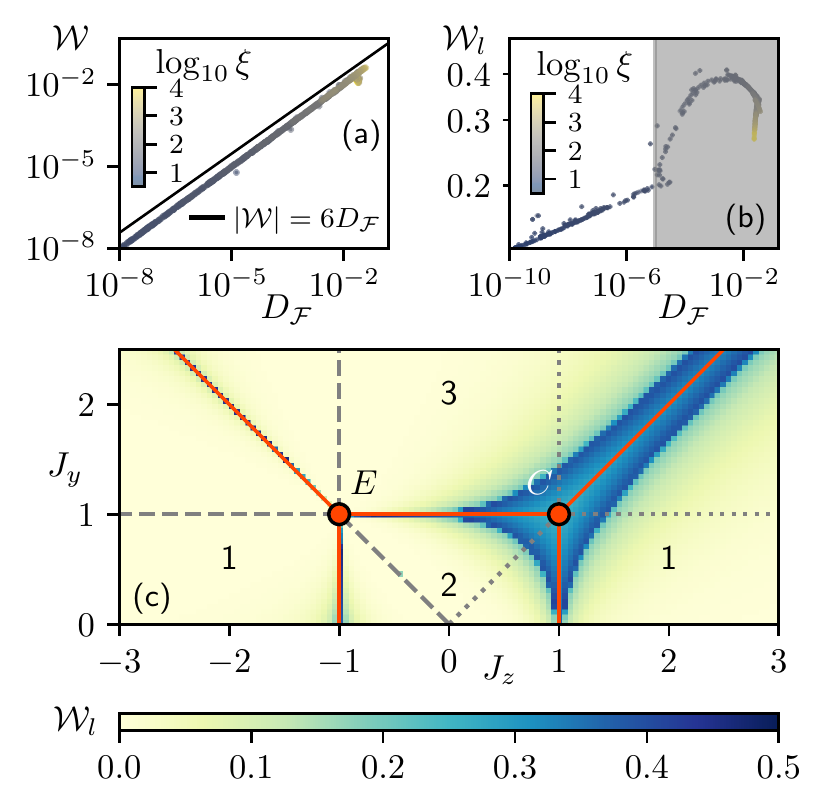}
  \caption{(a)-(b) Scatter plots comparing $|\mathcal{W}|$ in Eq.~(\ref{eq:w_es}) and  $|\mathcal{W}_l|$ in Eq.~(\ref{eq:wl}) with $D_\mathcal{F}$, for sizes $L{=}600$ and $L{=}400$, respectively. We see that $|\mathcal{W}|$ essentially coincides with $D_\mathcal{F}$, while $|\mathcal{W}_l|$ strongly correlates with $D_\mathcal{F}$ below the threshold $D_\mathcal{F}^*{\approx}10^{-9}$. The shaded area, $D_\mathcal{F}{>}D_\mathcal{F}^*$, corresponds to data points near critical regions with high correlation lengths, where the relationship between $D_\mathcal{F}$ and $\mathcal{W}_l$ breaks down. (c) $|\mathcal{W}_l|$  across the phase diagram of the XYZ model in Eq.~(\ref{eq:xyz_model}) for size $L{=}400$. When computing $|\mathcal{W}_l|$, we use a different Jordan-Wigner axis of quantisation for each region, labeled as follows: in $1$ we pick the $z$ quantisation axis, in $2$  we pick the $x$-axis,  and in $3$ the $y$-axis. 
  }
  \label{fig:xyz_wicks_scatter}
\end{figure}

Definition (\ref{eq:df}) allows us to determine the optimal free state $\sigma$ closest to $\rho$. Moreover, they are both diagonal in the same basis~\cite{turner_optimal_2017} that can be expressed in terms of the eigenoperators $a_j$ and $a_j^\dagger$. When $\rho$ corresponds to a Gaussian state, Wick's theorem dictates that $\langle a_i^\dagger a_i a_j^\dagger a_j\rangle_\rho = \langle a_i^\dagger a_i\rangle_\rho \langle a_j^\dagger a_j\rangle_\rho$, where $\langle a_i^\dagger a_i\rangle_\rho =\text{tr} (\rho a_i^\dagger a_i)$. However, if $\rho$ is non-Gaussian we do not expect this equality to hold any more. We thus define the Wick's theorem violation as
\begin{equation} \label{eq:w_es}
{\cal W}(\rho) = |\langle a_i^\dagger a_i a_j^\dagger a_j\rangle_\rho - \langle a_i^\dagger a_i\rangle_\rho \langle a_j^\dagger a_j\rangle_\rho|,
\end{equation}
which is a measure of how interacting a model is. In particular, ${\cal W}(\rho)$ can be calculated with the use of the dominant entanglement spectrum levels~\cite{SOM}. It is possible to show that 
\begin{equation}
{\cal W}(\rho) \leq \kappa D_{\cal F}(\rho),
\label{eqn:ineq}
\end{equation}
where $\kappa{=}6$ for the case of the $a_j$, $a_j^\dagger$ operators~\cite{SOM}. Hence, the interaction distance bounds from above the violation ${\cal W}$ of Wick's theorem. When applied to the case of the XYZ model we find that ${\cal W}$ and $D_{\cal F}$ almost coincide throughout the phase diagram, as shown in Fig.~\ref{fig:xyz_wicks_scatter}(a). 

The operators $a_i$, $a_i^\dagger$ are, in general, related to $c_j$, $c_j^\dagger$ of the underlying fermion lattice model (\ref{eq:fermion_model}) through a non-linear and non-local transformation. Hence, in order to determine~(\ref{eq:w_es}) experimentally, one needs full state tomography. As this is in general unrealistic to obtain in typical experiments, we apply the violation of Wick's theorem to the local operators $c_j$, $c_j^\dagger$. For convenience, we can employ the spin representation with quantisation axis taken to be the one for which the coefficient in the model is largest in absolute value, as shown in Fig.~\ref{fig:xyz_wicks_scatter}(c). For instance, where $|J_z| \geq |J_y|, |J_z|$ we define the violation of the local Wick's theorem as
\begin{eqnarray}\label{eq:wl}
\nonumber 	\mathcal{W}_{l}(\rho) &= & |\langle Z_{i} Z_{i+1} \rangle_\rho - \langle Z_{i} \rangle_\rho \langle Z_{i+1} \rangle_\rho  \\
&-& \langle Y_{i}X_{i+1} \rangle_\rho \langle X_{i}Y_{i+1} \rangle_\rho + \langle X_{i}X_{i+1} \rangle_\rho \langle Y_{i}Y_{i+1} \rangle_\rho|,\quad 
\end{eqnarray}
which is given in terms of two-spin correlators that are experimentally accessible. While ${\cal W}_l$ does not necessarily satisfy the inequality (\ref{eqn:ineq}), we determined numerically that it is tightly related to $D_{\cal F}$ with a monotonic one-to-one correspondence in the gapped region of the XYZ model, as shown in Fig.~\ref{fig:xyz_wicks_scatter}(b), where discrepancies from this behaviour only emerge near the critical regions, due to the finite-size effects. Thus, ${\cal W}_{l}$  can successfully identify the emerging freedom of the XYZ model. 
In Fig.~\ref{fig:xyz_wicks_scatter}(c), we evaluated ${\cal W}_{l}$ throughout the phase diagram of the XYZ model, finding very similar behaviour to  $D_{\cal F}$ in Fig.~\ref{fig:xyz_df_vne_pd}(b). Hence, the violation of the local Wick's theorem ${\cal W}_{l}$ provides the same information as $D_{\cal F}$, while it can in principle be measured in the laboratory.

\begin{figure}
\centering
\includegraphics[width=0.45\textwidth]{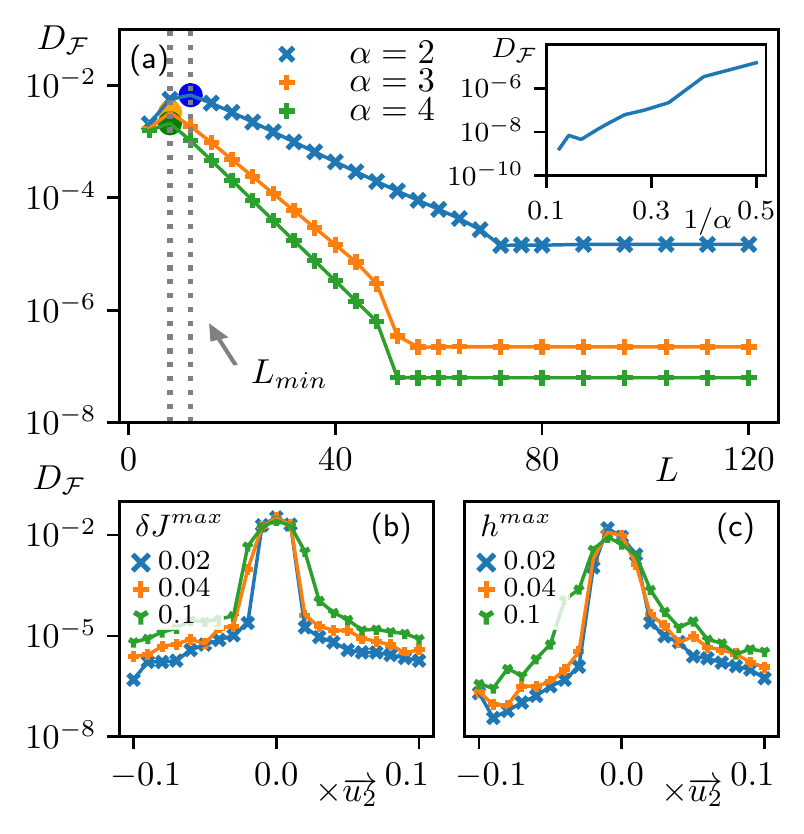}
\caption {
(a) $D_\mathcal{F}$ as a function of system size for the long-range XYZ model in Eq.~(\ref{eq:fermion_longrange_model}) with fixed $J_x{=}-1.0$, $J_y{=}-1.0$ and $J_z{=}5.0$ and various $\alpha$. Inset shows  the $L\rightarrow \infty$ saturation value of $D_f$ as a function of interaction range.  As expected, the saturation value becomes close to zero in the limit of short-range interactions.
(b) $D_\mathcal{F}$ across the $\ora{u_2}$ cut with different amounts of per-site randomness $\delta J^\mathrm{max}$ applied to the couplings $J_x$, $J_y$ and $J_z$ at system size $L{=}100$. (c) $D_\mathcal{F}$ across the $\ora{u_2}$ cut with a random local field of strength $h^\mathrm{max}$ applied to every site of a $L=100$ system.  }
\label{fig:RobustnessChecks}
\end{figure}

{\bf \em Robustness under realistic conditions.--} Finally, we 	consider the robustness of previous results when we move away from the exact XYZ model and introduce variations that model realistic experimental conditions. For example, in a cold atom implementation, the interactions between the constituent particles are characterised by a long-range algebraic decay~\cite{carr2009cold,mazurenko2017cold}. Moreover, there might be inhomogeneities in the engineered couplings due to imperfections in the laser control procedures or spurious random local potentials.

We first consider the effect a polynomial profile of interactions has on the behaviour of $D_{\cal F}$. We introduce a long-range XYZ model
\begin{align} \label{eq:fermion_longrange_model}
	\begin{split}
		H^\mathrm{LR} = \sum_{i,n} \frac{1}{|i-n|^\alpha}\left( J_{x} X_i X_{i+n} + J_{y} Y_i Y_{i+n} + J_z Z_i Z_{i+n}\right),
	\end{split}
\end{align}
where $\alpha$ controls the power-law decay of the couplings. The ground-state properties of this model can be captured using finite DMRG by expressing the algebraically decaying interaction as a sum of exponentials in order to represent the Hamiltonian as a matrix-product operator~\cite{pirvu2010matrix,o2018efficient}. In Fig.~\ref{fig:RobustnessChecks}(a) we show the behaviour of $D_{\cal F}$ as a function of system size in the long-range model. We picked a representative point which is in the gapped, antiferromagnetic phase for the entire range of $\alpha$ values considered~\cite{maghrebi2017continuous}. As in the short-range model, $D_{\cal F}$ decreases exponentially fast after the system exceeds a certain size $L_\mathrm{min}$. Note that, in contrast to the short-range case, $D_{\cal F}$ now levels off at a very small but non-zero value as $L{\to}\infty$, indicating that the model does not become completely free in the thermodynamic limit. The saturation value depends on the couplings and $\alpha$. Nevertheless, despite the saturation, we still observe an exponential reduction of $D_{\cal F}$ over several orders of magnitude, showing that the main characteristics of the model in the gapped phase remain similar when the range of interactions is changed drastically from ultra-local to polynomial range. 

A second type of robustness checks we performed  is the effect of experimental noise on $D_{\cal F}$. To model this we firstly introduce randomised couplings on each site.  In Fig.~\ref{fig:RobustnessChecks}(b) the couplings along the $\ora{u_2}$ cut are sampled uniformly from $[J_i - \delta J^\mathrm{max}, J_i + \delta J^\mathrm{max}]$ on each site with $D_{\cal F}$ remaining stable and increasing only a small amount up to large variations in the couplings. We additionally consider the impact of a spurious local magnetic field in the $z$-direction. In Fig.~\ref{fig:RobustnessChecks}(c) a random local field sampled uniformly from $[-h^\mathrm{max}, h^\mathrm{max}]$ was added on each site of the chain along the $\ora{u_2}$ cut. $D_{\cal F}$ also shows stability under this class of perturbations. Hence, the emerging freedom of the XYZ model persists in the presence of experimental imperfections that break the integrability of the XYZ model  in (\ref{eq:fermion_model}), while the behaviour of its ground-state correlations, as witnessed by $D_\mathcal{F}$ and Wick's theorem violation, remains largely the same.

{\bf \em Conclusions.--} There is a stark contrast between the behaviour of genuinely interacting systems and free ones in terms of their complexity in their description as well as their physical properties such as their thermalisation and out-of-equilibrium dynamics. Baxter has demonstrated that the XYZ model, which encompasses a large family of physically relevant models,  behaves in the thermodynamic limit as free, although it incorporates fermionic interactions. Here, we identified the system size conditions for the freedom to emerge near and away the critical regions of the model as a function of the correlation length of the system. 
 As our method does not rely on the integrability techniques, which are mainly restricted to one spatial dimension, it could be applied to other non-integrable 1D systems or even 2D models. Moreover, we quantified the emergent Gaussian behaviour in the XYZ model for the experimentally relevant cases of finite system sizes, long-range interaction potentials, as well as inhomogeneous couplings and random  local potentials. We proposed a way to observe the emergence of Gaussianity in the correlations of the XYZ model in terms of observables that can be directly measured in the laboratory. As Gaussianity emerges exponentially fast with system size, we anticipate that our findings can be experimentally verified in several experimental realisations of  XYZ-type models, both in solid state materials as well as synthetic ultracold atom systems~\cite{pinheiro2013x,pelegri2019quantum,tarruell2018quantum,jepsen2020spin,scheie2021detection,gring2012relaxation,murmann2015antiferromagnetic}.
 
{\bf \em Acknowledgements.--} We would like to thank Frank Verstraete and Chrysoula Vlachou for inspiring conversations. This work was supported by the EPSRC grant EP/R020612/1. Statement of compliance with EPSRC policy framework on research data: This publication is theoretical work that does not require supporting research data.

\bibliography{xyz_df}

\newpage{}
\clearpage{}
\onecolumngrid
\begin{center}
\textbf{\large Supplemental Online Material for ``Emergence of Gaussianity in the thermodynamic limit of interacting fermions'' }\\
\vspace*{0.2cm}
{\small Gabriel Matos, Andrew Hallam, Aydin Deger, Zlatko Papi\'c, and Jiannis K. Pachos
 }
\\
{\small \emph{School of Physics and Astronomy, University of Leeds, Leeds LS2 9JT, United Kingdom}}
\end{center}

{\small In this Supplementary Material, we show that the interaction distance bounds the Wick theorem violation. We also provide expressions for the correlation length in the XYZ model in the thermodynamic limit as a function of couplings, and we give some technical details about the DMRG simulations used in the main text.
}

\setcounter{equation}{0}
\setcounter{figure}{0}
\setcounter{table}{0}
\setcounter{page}{1}
\setcounter{section}{0}
\renewcommand{\theequation}{S\arabic{equation}}
\renewcommand{\theHequation}{S\arabic{equation}}
\renewcommand{\thefigure}{S\arabic{figure}}
\renewcommand{\theHfigure}{S\arabic{figure}}
\renewcommand{\thesection}{S\Roman{section}}
\renewcommand{\theHsection}{S\Roman{section}}
\renewcommand{\thepage}{S\arabic{page}}
\vspace{0.5cm}
\twocolumngrid
\section{Wick's theorem}

Consider the case of a free fermion system in its ground state $\ket{\psi_0}$. When the system is bipartitioned in $A$ and $B$ then its reduced density matrix $\rho =\text{tr}_B(\ket{\psi_0}\bra{\psi_0})$ can be expressed of a thermal state $\rho = e^{-H_E}$ where $H_E$ is the entanglement Hamiltonian. As the initial model is free the entanglement Hamiltonian is also free and its correlations can be given in terms of its fermionic eigenoperators $a_j$, $a_j^\dagger$ as
\begin{equation}
H_E = \sum_j \epsilon_j a^\dagger_j a_j,
\label{eqn:entHam}
\end{equation}
where $\epsilon_j$ are the single particle energies. From this density matrix we can calculate the two-point correlator as
\begin{equation}
\langle a_i^\dagger a_i\rangle_\rho = {1 \over e^{\epsilon_i}+1}.
\label{eqn:2point}
\end{equation}
Wick's theorem provides the means to calculate higher-point correlators in terms of two-point correlators of such free models. In the case of the four-point operator $a_i^\dagger a_i a_j^\dagger a_j $ in terms of the ground state $\rho$ the Wick's theorem takes the form
\begin{equation}
\langle a_i^\dagger a_i a_j^\dagger a_j \rangle_\rho = \langle a_i^\dagger a_i \rangle_\rho \langle a_j^\dagger a_j \rangle_\rho,
\label{eqn:Wicks1}
\end{equation}
where $\langle {\cal O}\rangle_\rho =\text{tr} (\rho{\cal O})$. 

In the case of an interacting fermions system the Wick's theorem cannot be applied any more. Instead, we can quantify the violation of the condition (\ref{eqn:Wicks1}) in terms of
\begin{equation}
{\cal W}(\rho) = |\langle a_i^\dagger a_i a_j^\dagger a_j \rangle_\rho - \langle a_i^\dagger a_i \rangle_\rho \langle a_j^\dagger a_j \rangle_\rho|
\label{eqn:W}
\end{equation}
We can use ${\cal W}$ to quantify the effect of interactions in the four point correlations in a similar way we use the interaction distance, $D_{\cal F}(\rho)$. In particular, we can show that ${\cal W}(\rho)$ is upper bounded by $D_{\cal F}(\rho)$. The interaction distance also provides the optimal free model $\sigma$ that is closest to $\rho$. It is clear that ${\cal W}(\sigma)=0$ as $\sigma$ is a free state. We can then write 
\begin{eqnarray}
{\cal W}(\rho) = |\langle a_i^\dagger a_i a_j^\dagger a_j \rangle_\rho - \langle a_i^\dagger a_i a_j^\dagger a_j \rangle_\sigma -\nonumber \\ \langle a_i^\dagger a_i \rangle_\rho \langle a_j^\dagger a_j \rangle_\rho + \langle a_i^\dagger a_i \rangle_\sigma \langle a_j^\dagger a_j \rangle_\sigma| \leq
\nonumber \\
|\text{tr}(a_i^\dagger a_i a_j^\dagger a_j(\rho-\sigma))| + 
\nonumber \\
|\langle a_i^\dagger a_i \rangle_\rho| |\text{tr} (a_j^\dagger a_j (\rho-\sigma))| +
\nonumber \\
|\text{tr}(a_i^\dagger a_i (\rho-\sigma))| |\langle a_j^\dagger a_j\rangle_\sigma|.
\label{eqn:ineq1}
\end{eqnarray}
We have that 
\begin{equation}
\text{tr}({\cal O} (\rho-\sigma)) \leq {2\|{\cal O}\| } D_{\cal F}(\rho),
\end{equation}
where $ \|{\cal O}\|$ can be taken to be the largest eigenvalue of the operator ${\cal O}$ \cite{patrick2019efficiency}. Hence, from (\ref{eqn:ineq1}) we obtain
\begin{equation}
{\cal W}(\rho) \leq {6} D_{\cal F}(\rho),
\end{equation}
i.e. the violation of the Wick's theorem due to the presence of interactions is bounded from above by the interaction distance. 

For an interacting fermionic system we assume the form of the interacting entanglement Hamiltonian to be
\begin{equation}
H^\text{int}_E = \sum_i \epsilon_i a^\dagger_i a_i +  \sum_{i,j} \epsilon_{ij} a^\dagger_i a_ia^\dagger_j a_j + \dotsm,
\label{eqn:entHamint1}
\end{equation}
where $\epsilon_i$ are the single particle energies and $\epsilon_{ij}$ are the two particle energies of the entanglement Hamiltonian. This Hamiltonian is diagonal in the basis of the eigenoperators $a_i$, $a_i^\dagger$. The two-point correlator with respect to 
\begin{equation}
\rho = \exp(-H^\text{int}_E). 
\end{equation}
is also given by (\ref{eqn:2point}). On the other hand, the four-point correlator is now given by
\begin{equation}
\langle a_i^\dagger a_ia_j^\dagger a_j\rangle_\rho = {1 \over e^{\epsilon_{ij}} + e^{\epsilon_{ij}-\epsilon_i} +e^{\epsilon_{ij}-\epsilon_j}  + 1}.
\end{equation}
With the help of these expressions we can calculate the violation (\ref{eqn:W}) of the Wick's theorem, due to interactions, exclusively in terms of the entanglement energies $\epsilon_i$, $\epsilon_j$ and $\epsilon_{ij}$. In practice, we determine these entanglement energies in the following way. We assume that the first two smallest levels, excluding the one corresponding to the normalisation, correspond to the two smallest energies $\epsilon_i$ and $\epsilon_j$. We then assume that the level closest to their sum corresponds to $\epsilon_{ij}$.

\section{Correlation length for the XYZ model}

In a seminal paper~\cite{Sutherland1970}, Sutherland showed that the following XYZ model
\begin{equation}
H = - \sum_{i=1}^L X_i X_{i+1} + J_y Y_i Y_{i+1} + J_z Z_i Z_{i+1},
\end{equation}
is closely related to the zero-field eight-vertex model. Using the symmetries of both models, Baxter~\cite{ Baxter1971b} introduced a rearrangement procedure and parameterisation of XYZ coefficients $J_y$ and $J_z$ in terms of elliptic functions. In the following, we use the same notation and definition of $\Gamma$ and $\Delta$ as described in Baxter's book~\cite{Baxter1985}. The parametrisation of these quantities in the principal regime $|\Gamma_r|=|J_y|\leq 1$ and $\Delta_r=J_z\leq -1$ is given by
\begin{align*}
	\Gamma_r &=(1+k~ \mathrm{sn}^2 i \lambda )/(1-k ~ \mathrm{sn}^2 i\lambda),\\
	\Delta_r &=-\mathrm{cn}~i \lambda~ \mathrm{dn}~i \lambda /(1-k ~ \mathrm{sn}^2i\lambda),
\end{align*}
where $\mathrm{sn},\mathrm{cn},\mathrm{dn}$ are the Jacobian elliptic functions, $k$ and $\lambda$ are elliptic function parameters whose natural domains are as follows
\begin{align*}
	0 &\leq k \leq 1,\\
	0 &\leq \lambda \leq \mathcal{I}(k'),
\end{align*}
where $\mathcal{I}(k)$ denotes the complete elliptic integral of the first kind and the complementary modulus is defined by $k' \equiv \sqrt{1-k^2}$ with modulus $k$. We also set $\mu\equiv \pi \lambda/\mathcal{I}(k')$ and $x=e^{-\pi \lambda / 2 \mathcal{I}(k)}$ for later convenience.

We note that one can obtain results for the entire phase diagram by using a rearrangement procedure, i.e., permuting the $J_y$ and $J_z$ terms by a spin rotation, which introduces artificial discontinuities between different regions. The same result can also be obtained using a more elegant prescription based on the modular properties of the elliptic functions~\cite{Ercolessi2013}.

Using this parametrisation, Johnson and Baxter~\cite{Johnson1973, Baxter1985} obtained the analytical formula for the correlation length $\xi$ of the XYZ model in the ordered regime as
\[ \xi^{-1}= 
\begin{cases}
    -\ln k_1 \quad & (\mu\leq\pi/2) ,\\
    -\ln \frac{k_1}{\mathrm{dn}^2\left[\frac{\mathcal{I}(k_1) \mathcal{I}(k')}{2\mathcal{I}(k)}-\mathcal{I}(k_1'),k_1'\right]} \quad & (\pi/2<\mu) ,
    \end{cases}\]
where $k_1$ is the elliptic modulus with nome $x^2$. We note that negating $J_z$ term yields the XYZ Hamiltonian considered in the main text.

\section{Methods}
\label{sec:methods}

When performing DMRG to find the ground states of the XYZ model in the main text, we found that the exponentially vanishing quasi-degeneracy with system size was resolved by simply picking an appropriate, physically motivated, initial state. This amounted to picking a state in the basis for which the operator with the highest coefficient (in absolute value) in the Hamiltonian is diagonal. The state picked depends on the sign of this largest coefficient; if it is negative, a Néel state $\uparrow \downarrow ... \uparrow \downarrow$ in the appropriate basis is picked; if it is positive, a fully polarised state $\uparrow \uparrow ... \uparrow$ in the appropriate basis is picked. For instance if $\abs{J_y} > \abs{J_x}$, $\abs{J_y} > \abs{J_z}$ and $J_y < 0$, then a $Y$-polarized Néel initial state is picked. 

The DMRG code was run with bond dimension $\chi = 128$ for the long-range and random fields data in Figure~\ref{fig:RobustnessChecks}. For the data in  Figure~\ref{fig:xyz_df_vne_pd} and Figure~\ref{fig:xyz_wicks_scatter}(a), the bond dimension was allowed to scale as necessary; for Figure~\ref{fig:xyz_wicks_scatter}(b), (c), bond dimension 128 was used, and bond dimension 512 was used for Figure~\ref{fig:xyz_scalings}.

\end{document}